\documentclass[aps,prl,twocolumn,groupedaddress]{revtex4}

\usepackage{graphicx}
\usepackage{amsmath,amssymb}

\begin{document}

\title{Magnetoelastic nonlinear metamaterials}

\author{Mikhail Lapine}
\email[ Corresponding author: ]{ mlapine@uos.de }
\affiliation{Nonlinear Physics Centre, Research School of Physics and Engineering, Australian National University, ACT 0200, Australia}

\author{Ilya V. Shadrivov}
\affiliation{Nonlinear Physics Centre, Research School of Physics and Engineering, Australian National University, ACT 0200, Australia}

\author{David A. Powell}
\affiliation{Nonlinear Physics Centre, Research School of Physics and Engineering, Australian National University, ACT 0200, Australia}

\author{Yuri S. Kivshar}
\affiliation{Nonlinear Physics Centre, Research School of Physics and Engineering, Australian National University, ACT 0200, Australia}



\begin{abstract}
We introduce the concept of magnetoelastic metamaterials with electromagnetic properties depending on elastic deformation.
We predict a strong nonlinear and bistable response of such metamaterials caused by
their structural reshaping in response to the applied electromagnetic field.
In addition, we demonstrate experimentally the feasibility of the predicted effect.
\end{abstract}


\maketitle


The study of metamaterials became a prominent area of research, bringing
together theoretical and applied electrodynamics, as well as electrical engineering
and material science. Being initially intended to achieve negative refraction~\cite{SPV0,PenPW},
metamaterials quickly covered a much wider range of applications from microwaves to optics.
Nonlinear metamaterials~\cite{ZSK3,LGR3} established a new research direction
\cite{ASB4,OMR4,SSA5,PopSha6,PadTayHig6,GabIndLit6,SymSolYou8,PouHuaSmi10}
giving rise to fruitful ideas for tunable and active artificial materials
\cite{BoaGriKiv11}.

The initial way to provide strong nonlinearity to metamaterials was found in either
employing a nonlinear host medium \cite{ZSK3} or by engineering the elements of
a metamaterial with a nonlinear component \cite{LGR3}.
In those approaches the nonlinear response is obtained on the level of individual elements.
On the other hand, by varying
the mutual interaction between elements one can efficiently control bulk metamaterial
properties \cite{LapPowGor09,PowLapGor10}.
It is therefore quite promising to explore the possibilities
of nonlinear mutual interaction in metamaterials.

In this Letter, we propose a novel concept of \emph{magneto\-elastic metamaterials},
where nonlinearity arises from a collective response.
This is achieved by providing a mechanical degree of freedom so that the
electromagnetic interaction in the metamaterial lattice is supplemented by elastic interaction.
This enables the electromagnetically induced forces to change the metamaterial shape,
thus changing its effective properties. Consequently, such a metamaterial exhibits efficient
self-action, which leads to strong nonlinear effects and non-trivial bistability.

\begin{figure}
\centering
\includegraphics[width=1\columnwidth]{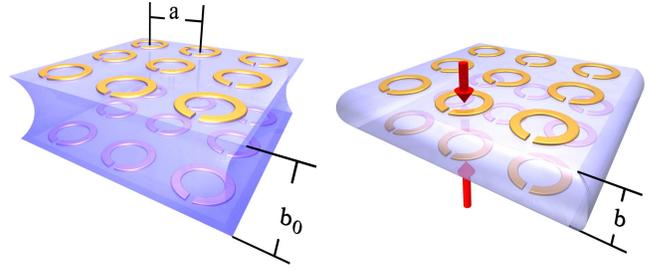}
\caption{\label{F1} (Color online) Schematic of anisotropic magnetic metamaterial
(two layers are shown) assembled with an elastic medium.
Left: metamaterial before the electromagnetic field is applied.
Right: metamaterial is compressed by the electromagnetic forces acting between the elements.
}
\end{figure}

To illustrate this concept, we consider a magnetic metamaterial composed of a lattice of resonant elements,
such as split-ring resonators (SRRs) or capacitively-loaded rings (CLRs).
We select an anisotropic arrangement (Fig.~\ref{F1}) with all the resonators having the same
orientation, so that the inter-layer distance $b$ can be made sufficiently small to ensure a
stronger interaction. For simplicity, we choose a circular shape of resonators, however this
does not limit the generality of the predicted phenomena. For the convenience
of analytical expressions, we use the dimensionless lattice parameters $a$ and $b$, normalized
to the resonator radius $r_0$.

In response to electromagnetic waves with magnetic field $H_0$ along the axial direction,
such a metamaterial shows resonant magnetic behavior \cite{GLS2}.
The currents induced in the resonators not only affect each other through mutual
inductance, but also result in an attractive force between the resonators
(it is attractive provided that the neighboring currents are in phase).
So if the resonators are allowed to move along the axial direction,
they will displace from their original positions, thus changing their mutual impedance,
which in turn affects the current amplitudes, interaction forces, and so on.
The balance is kept with a restoring force, which originates from
the elastic properties of the host medium.

We consider a quasi-stationary regime, where
the characteristic response times for mechanical movements are much
larger than the period of electromagnetic oscillations.
We also assume that the electromagnetic resonance frequency
is such that the wavelength is much larger than element
size and lattice constants (this can be easily achieved e.g.\ by using
CLRs with an appropriate capacitance, or broadside-coupled SRRs
\cite{MarMarSor} with sufficiently small gaps).
In this case, the currents induced in neighboring elements are
all in phase and also uniform along the resonator circumference,
so that the attraction force can be evaluated
in the same way as the static force for DC currents, weighted with
a time-averaged current amplitude, $I/\sqrt{2}$.
At the same time, electric interaction is irrelevant
in a quasi-static situation, or otherwise can be minimized by choosing
an appropriate mutual orientation of the rings \cite{PowLapGor10}.

For identical rings sharing a common axis and carrying currents with amplitude $I$,
the force acting between them has only the axial component, calculated as \cite{Landau}
\begin{equation}
F_\text{i} = \frac{\mu_0 I^2}{2 \sqrt{4 + b^2}} \:
    \left( \mathcal{E} \: \frac{2 + b^2}{b^2} - \mathcal{K} \right),
\label{fzu}
\end{equation}
where $\mathcal{E}$,  $\mathcal{K}$ are the complete elliptic integrals
of first and second kind with the parameter $\varkappa^2 = {4}/{(4 + b^2)}$.

In the limit of remote rings ($b \gg 1$), the expression \eqref{fzu} can be
expanded with respect to $1/b$ as a small parameter, with elliptic integrals
evaluated explicitly. This yields a long distance limit
\begin{equation}
\lim_{b \rightarrow \infty} F_\text{i} = \frac{\mu_0 I^2}{2} \: \frac{3 \pi}{2} \: \left( \frac{1}{b} \right)^4.
\label{ffar}
\end{equation}
Naturally, this result agrees, up to different $O\bigl( 1 / b^6 \bigr)$ terms,
with the dipole approximation valid for large distances.
This allows us to take into account far-neighbor interactions consistently: the summation over remote
rings quickly converges thanks to the fourth power decay, so within a relevant volume
we can still assume that the currents are not affected by retardation.

We assume that the lateral lattice constant $a$ is reasonably large
so that additional forces between the rings in the neighboring columns can be neglected.
Then the total compression force acting between any two rings in the bulk,
can be shown to have the following form:
\begin{equation}
F_\text{I} (b)
\approx \sum_{n=1}^{N } n \cdot F_\text{i} (n b) + S_N,
\label{ftot}
\end{equation}
where $N$ is number of rings for which the approximate solution \eqref{ffar} is not yet
sufficiently precise, while for the remaining rings the quickly converging summation over \eqref{ffar}
yields a minor addition $S_N$ which can be neglected in practice.
The value of $N$ depends on the lattice density, being for example
of the order of 100 for $b=0.1$ and of 20 for $b=1$.
For practical purposes, it is convenient to introduce a specific
multiplication factor $\beta$ to obtain $F_\text{I} = \beta F_\text{i}$ directly from \eqref{fzu}.
Empirically, it turns out that $\beta$ is closely proportional to $1/b$
so the exact summation \eqref{ftot} can be replaced with quick approximate calculation
\begin{equation}
F_\text{I} \approx \frac{\pi}{2} \: \frac{1}{b} \: F_\text{i}.
\label{fappr}
\end{equation}

\begin{figure}
\centering
\raisebox{3mm}{\includegraphics[width=0.99\columnwidth]{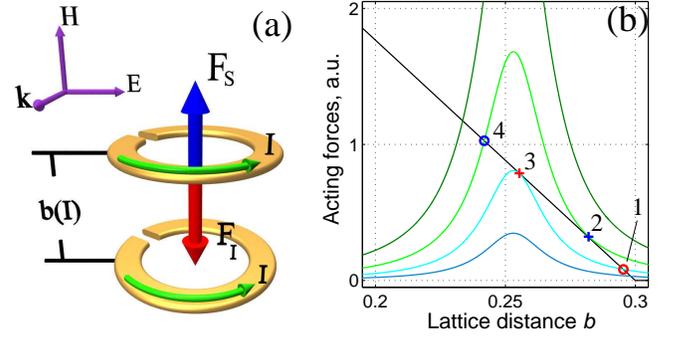}}
\caption{\label{F2} (Color online) (a)~Schematic of the forces acting on a ring within a metamaterial,
where the total compressing force resulting from current attraction, $F_\text{I}$,
is countered by the elastic force $F_\text{S}$, both being dependent
on the lattice distance $b$ which varies with the current amplitude;
(b)~An example of force magnitudes depending on the lattice distance,
where attraction forces $F_\text{I}$ for several current amplitudes are shown with
colored peaks and the counter-acting spring force $F_\text{S}$
with a black straight line. Stable equilibrium points are shown with circles while unstable ones with crosses.}
\end{figure}

We now assume that the elastic force which balances the magnetic attraction, obeys
a Hooke's law, $F_\text{S} (b) = k r_0 (b - b_0)$, with a generic stiffness coefficient $k$,
and the initial lattice constant $b_0$.
Thus the distance $b$ where equilibrium is achieved
will be determined by a balance between
the total compression force in the bulk $F_\text{I}$,
and the elastic force $F_\text{S}$ [Fig.~\ref{F2}(a)]:
\begin{equation}
\beta(b) \cdot F_\text{i} (b,I) + k r_0 (b - b_0) = 0.
\label{baleq}
\end{equation}
We must impose an artificial limit $b_{\text{min}}$ on how small $b$ can be,
to reflect the unavoidable technological restrictions as well as to
ensure a reasonable limit on the linear elasticity law, which simplifies the analysis.

Self-action, resulting in nonlinear behavior, occurs through the mutual
inductance between the rings, which also depends on $b$.
In metamaterials, the effect of mutual interaction between all the rings,
is accounted for by the so-called lattice sum $\Sigma$, which depends
on the lattice type and parameters. It can be numerically
calculated as explained in Ref.~\cite{GLS2}.
The complete impedance equation
\begin{equation}
\left[ Z + \text{i} \omega \mu_0 r_0 \Sigma (a, b) \right] \cdot I
    = - \text{i} \omega \pi r_0^2 \mu_0 H_0
\label{impeq}
\end{equation}
together with the equilibrium equation \eqref{baleq}, forms a system
of coupled equations, which can be numerically solved to yield
$b$ and $I$ for a given incident amplitude $H_0$ and frequency $\omega$.
Then the magnetization of the metamaterial is obtained as
$M = I \nu \pi r_0^2 = (\pi/r_0 a^2)(I/b)$,
where the effect is further enhanced through the dependence of the volumetric
density of rings, $\nu = 1/(r_0 a^2 b)$, on $b$.
Thus, we can characterize our metamaterial with
a nonlinear and resonant $M(H_0,\omega)$ dependence.

To outline the expected phenomena, we depict some examples of the interplay of
the involved forces in Fig.~\ref{F2}.
The solutions to the balance \eqref{baleq} are graphically seen as the crossing
points, and the stable equilibrium positions are
such that $F_\text{I} < F_\text{S}$ for an attempted decrease in~$b$.
Note that the resonant nature of the currents, induced in the rings
depending on $b$ [see Eq.~\eqref{impeq}], defines the resonant character
of the force $F_\text{I}$.
Thus, when three mathematical solutions are available, only two are actually stable;
or otherwise there is a single stable state.
The phenomenology is qualitatively clear:
when the current amplitude exceeds certain threshold
[where indicated with cross ``2'' in Fig.~\ref{F2}(b)],
the initial ``right-side'' equilibrium (such as at circle ``1'')
cannot be achieved, so the lattice distance $b$ attempts to collapse.
However this also changes the mutual interaction dramatically, leading to
a significant shift of the resonance frequency, so the current magnitude
drops, permitting the other (``left-side'') equilibrium state (Fig.~\ref{F2}, circle ``4''),
corresponding to the same force curve.
On the other hand, with a decreasing amplitude, the ``left-side'' balance remains stable as long as
the peak attraction force is sufficient to counter the elastic force
(down to an unstable point, cross ``3''), from where the system jumps
back to the corresponding ``right-side'' solution (circle ``1'').

\begin{figure}
\centering
\includegraphics[width=0.99\columnwidth]{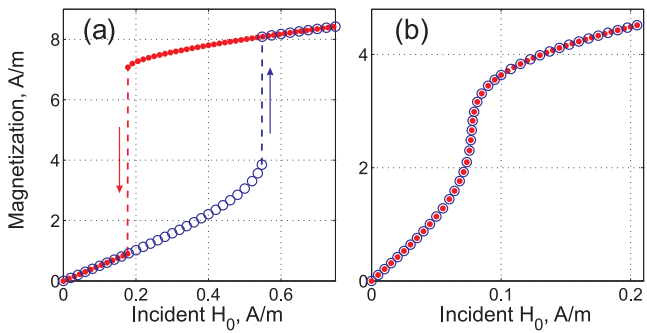}
\caption{\label{F3} (Color online) Magnetization $M(H_0,\omega)$
in metamaterial vs. incident amplitude $H_0$, observed at relative
frequencies of 0.55 (a) or 0.6 (b),
for increasing (blue circles) and decreasing (red bullets) amplitudes.}
\end{figure}

We illustrate typical patterns of the arising nonlinearity (Fig.~\ref{F3})
with elements of radius $r_0 = 5$\,mm, resonating
individually at 1\,GHz (the frequency values below are
normalized with respect to the corresponding angular frequency $\omega_0$),
with a quality factor of 100.
These are arranged in a metamaterial with
$a = 4$, $b_0 = 0.3$, $b_{\text{min}} = 0.1$, with a stiffness
coefficient $k = 0.44$\,mN/m.
As we show below, although the required coefficient is rather small with respect
to bulk conventional materials, it can be realized in practice with the help
of appropriately bend thin filaments or small plastic springs with realistic
geometrical parameters, which fit to the suggested geometry.

\begin{figure}[b]
\centering
\includegraphics[width=0.99\columnwidth]{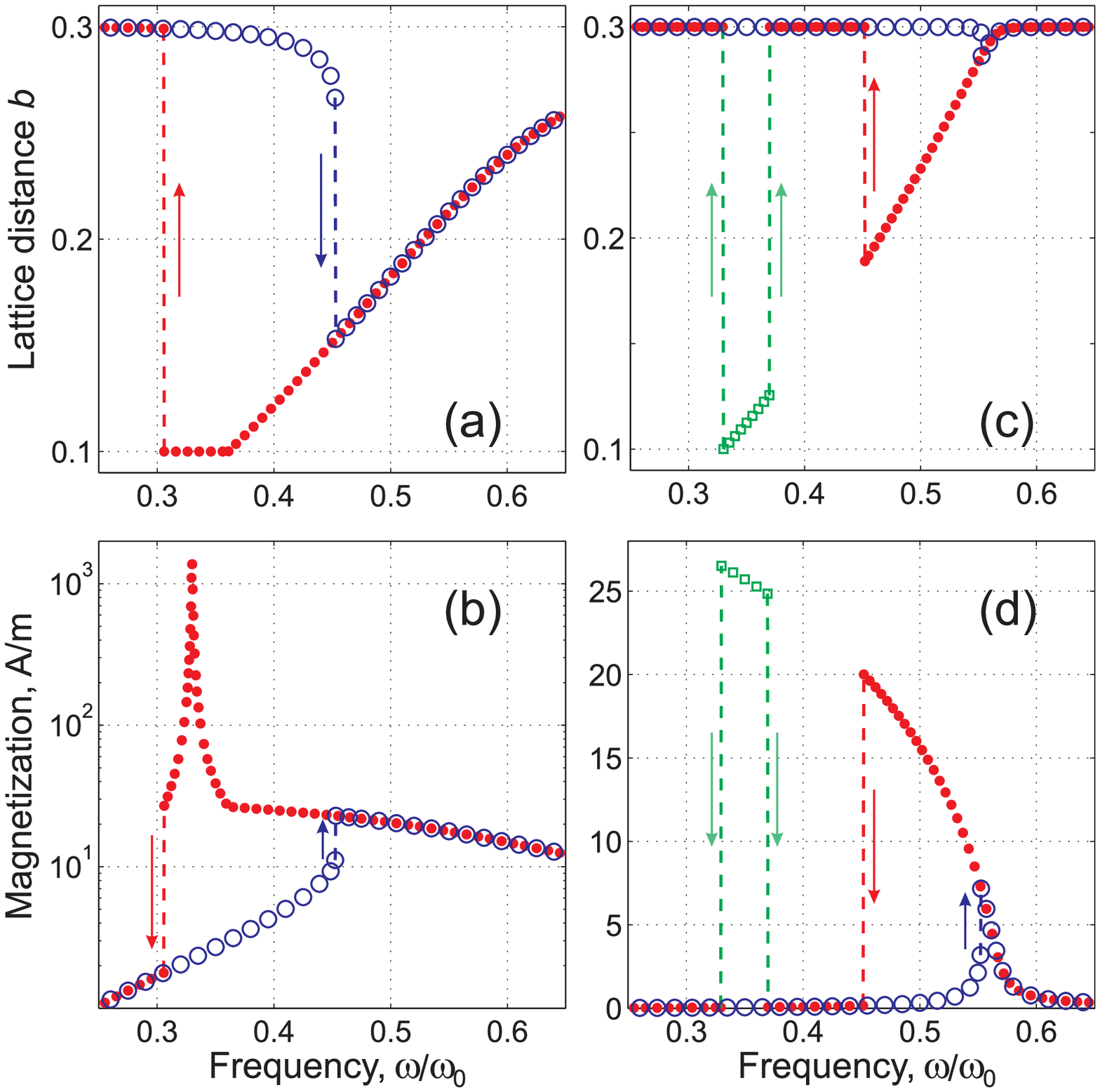}
\caption{\label{F4} (Color online) Increasing (blue circles) and decreasing (red bullets)
frequency dependence of (a,\,c) the lattice distance $b$
and (b,\,d) magnetization $M(H_0,\omega)$ observed at incident
amplitudes $H_0$ of (a,\,b) 20\,A/m or (c,\,d) 0.4\,A/m.
The quasi-inaccessible stable conformation is shown with green squares.
Note the logarithmic vertical scale in panel (b).}
\end{figure}

At frequencies lower than the eigenfrequency of the initial state,
we observe a slightly nonlinear $M(H_0)$ dependence as the amplitude grows,
until the metamaterial abruptly switches to a stronger compression.
However, when the amplitude is decreased, the metamaterial remains in the
compressed state until much lower magnitudes, exhibiting
a hysteresis-like behavior [Fig.~\ref{F3}(a)].
But close to the original resonance, the hysteresis disappears while the nonlinearity
is quite strong [Fig.~\ref{F3}(b)].

More spectacular phenomena can be observed with the frequency dependence, which reveals
the entire drama of complex bistable behavior (Fig.~\ref{F4}).
With moderate to high amplitudes, lattice distance $b$ declines slowly with
growing frequency, until at some stage the initial balance of forces is lost
and metamaterial jumps to a more compressed state, from where it
gradually returns back to the original state with further frequency increase;
the magnetization pattern reflects these changes
[see the blue circles in Figs.~\ref{F4}\,(a,\,b)].
But when the frequency is decreased from the high values, the structure
remains in the compressed state across that threshold, and continues to compress
until the mechanical limit at $b_\text{min}$ is reached, where it remains until the currents induced
at still decreasing frequency become low enough to release the entire jump back
to the ground state [Fig.~\ref{F4}(a), red bullets]. This is followed by
the magnetization, but note that in the frequency range of full compression
the response of the metamaterial is linear and hence we can observe a purely linear
resonance [see the red bullets in Fig.~\ref{F4}(b)].

\begin{figure}
\centering
\includegraphics[width=0.99\columnwidth]{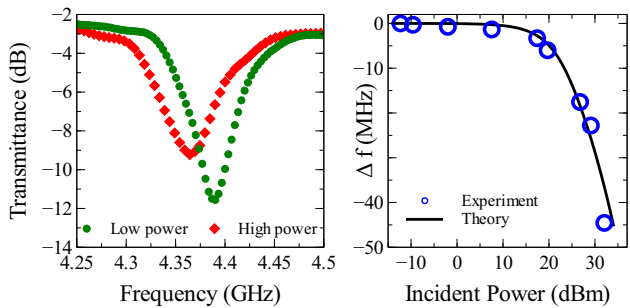}
\caption{\label{F5} (Color online)
Experimental observation of the magnetoelastic nonlinearity in a system of two resonators.
Left: measured transmission spectra at low ($-12.3$\,dBm) and high ($29$\,dBm) power.
Right: dependence of the resonance frequency on the incident power,
showing the experimental (circles) and theoretical (solid line) results.
}
\end{figure}

However, at low amplitudes, we can observe highly unusual behavior
[Figs.~\ref{F4}\,(c,\,d)]. The frequency hysteresis described above still applies,
but in addition there appears a frequency range where the metamaterial is
stable in a compressed state, but which, however, cannot be directly accessed
in this hysteresis loop [see the green squares in Figs.~\ref{F4}\,(c,\,d)].
The only ways to reach this range of compressions are to push the structure
once with an external mechanical force, or to temporarily increase the intensity.
Once there, the metamaterial remains stable in the corresponding frequency window, jumping
to the ground state when decreasing or increasing the frequency past the window limits.
This spectacular effect exists because below certain amplitudes, the currents
induced in the intermediate range of $b$ are not sufficient to hold the force
balance even at resonance; however they are still able to do it for smaller $b$
because of the effectively $1/b^2$ law in Eq.~\eqref{fappr}.
The range of amplitudes where this can be observed, is relatively narrow
(in our example, between 0.38 and 0.41~A/m),
with the corresponding frequency range becoming increasingly narrow with
decreasing $H_0$, until finally disappearing while the main hysteresis
is still in place up to very low amplitudes.

In order to demonstrate the plausibility of the predicted effect,
we perform a pump-probe experiment for a pair of closely spaced SRRs,
having 4.6\,mm radius, 1\,mm gap and made of 0.18\,mm-thick copper wire.
We suspend the SRRs parallel to the axis of a WR~229 rectangular waveguide,
at a distance of 1.4\,mm ($b_0 = 0.3 r_0$), on a dielectric rod with two groves,
so that the rings are able to swing towards each other when an attractive force is induced.
To provide for an elastic force, we use four U-shaped keratin filaments between the resonators.
The waveguide is excited by combined signal of the vector network analyzer (Rhode and Schwartz ZVB-20)
and a continuous wave pump generated by a signal generator (HP~8673B), which is amplified by a signal
amplifier (HP~83020A). Frequency scans are performed using a vector network analyser with a signal
power of $-30$\,dBm, while the continuous wave pump is applied at 4.38\,GHz.
The spectrum of the incident signal is measured using a broadband directional coupler.

Figure~\ref{F5} demonstrates a remarkable effect of increasing electromagnetic wave power,
with a resonance frequency shift of 44.5 MHz.
The experimental power dependence of the resonance matches very well with the theory developed above
(upon the straightforward amendments required to account for two resonators instead of a bulk material),
with the corresponding parameters and assuming a stiffness coefficient of about 1.3\,mN/m.
Note the latter is very close to the one used for theoretical results and therefore proves
them absolutely feasible.

In summary, we have proposed a novel type of metamaterial introducing mechanical degrees of
freedom, and demonstrated that this provides a clearly measurable effect with the
parameters very close to our assumptions.
The magnetoelastic coupling mechanism leads to many interesting nonlinear effects
which can be useful for further theoretical development as well as for future applications
in microwave, THz and optical range.


This work was supported by the Australian Research Council.
The authors thank M. Gorkunov and A. Sukhorukov for useful discussions.




\begin{thebibliography}{18}
\expandafter\ifx\csname natexlab\endcsname\relax\def\natexlab#1{#1}\fi
\expandafter\ifx\csname bibnamefont\endcsname\relax
  \def\bibnamefont#1{#1}\fi
\expandafter\ifx\csname bibfnamefont\endcsname\relax
  \def\bibfnamefont#1{#1}\fi
\expandafter\ifx\csname citenamefont\endcsname\relax
  \def\citenamefont#1{#1}\fi
\expandafter\ifx\csname url\endcsname\relax
  \def\url#1{\texttt{#1}}\fi
\expandafter\ifx\csname urlprefix\endcsname\relax\def\urlprefix{URL }\fi
\providecommand{\bibinfo}[2]{#2}
\providecommand{\eprint}[2][]{\url{#2}}

\bibitem[{\citenamefont{Smith et~al.}(2000)\citenamefont{Smith, Padilla, Vier,
  Nemat-Nasser, and Schultz}}]{SPV0}
\bibinfo{author}{\bibfnamefont{D.~R.} \bibnamefont{Smith}},
  \bibinfo{author}{\bibfnamefont{W.~J.} \bibnamefont{Padilla}},
  \bibinfo{author}{\bibfnamefont{D.~C.} \bibnamefont{Vier}},
  \bibinfo{author}{\bibfnamefont{S.~C.} \bibnamefont{Nemat-Nasser}},
  \bibnamefont{and} \bibinfo{author}{\bibfnamefont{S.}~\bibnamefont{Schultz}},
  \bibinfo{journal}{Phys.\ Rev.\ Lett.} \textbf{\bibinfo{volume}{84}},
  \bibinfo{pages}{4184} (\bibinfo{year}{2000}).

\bibitem[{\citenamefont{Pendry}(2001)}]{PenPW}
\bibinfo{author}{\bibfnamefont{J.~B.} \bibnamefont{Pendry}},
  \bibinfo{journal}{Physics World} \textbf{\bibinfo{volume}{14}},
  \bibinfo{pages}{47} (\bibinfo{year}{2001}).

\bibitem[{\citenamefont{Zharov et~al.}(2003)\citenamefont{Zharov, Shadrivov,
  and Kivshar}}]{ZSK3}
\bibinfo{author}{\bibfnamefont{A.~A.} \bibnamefont{Zharov}},
  \bibinfo{author}{\bibfnamefont{I.~V.} \bibnamefont{Shadrivov}},
  \bibnamefont{and} \bibinfo{author}{\bibfnamefont{Yu.~S.}
  \bibnamefont{Kivshar}}, \bibinfo{journal}{Phys.\ Rev.\ Lett.}
  \textbf{\bibinfo{volume}{91}}, \bibinfo{pages}{037401}
  (\bibinfo{year}{2003}).

\bibitem[{\citenamefont{Lapine et~al.}(2003)\citenamefont{Lapine, Gorkunov, and
  Ringhofer}}]{LGR3}
\bibinfo{author}{\bibfnamefont{M.}~\bibnamefont{Lapine}},
  \bibinfo{author}{\bibfnamefont{M.}~\bibnamefont{Gorkunov}}, \bibnamefont{and}
  \bibinfo{author}{\bibfnamefont{K.~H.} \bibnamefont{Ringhofer}},
  \bibinfo{journal}{Phys.\ Rev.~E} \textbf{\bibinfo{volume}{67}},
  \bibinfo{pages}{065601} (\bibinfo{year}{2003}).

\bibitem[{\citenamefont{Agranovich et~al.}(2004)\citenamefont{Agranovich, Shen,
  Baughman, and Zakhidov}}]{ASB4}
\bibinfo{author}{\bibfnamefont{V.~M.} \bibnamefont{Agranovich}},
  \bibinfo{author}{\bibfnamefont{Y.~R.} \bibnamefont{Shen}},
  \bibinfo{author}{\bibfnamefont{R.~H.} \bibnamefont{Baughman}},
  \bibnamefont{and} \bibinfo{author}{\bibfnamefont{A.~A.}
  \bibnamefont{Zakhidov}}, \bibinfo{journal}{Phys.\ Rev.~B}
  \textbf{\bibinfo{volume}{69}}, \bibinfo{pages}{165112}
  (\bibinfo{year}{2004}).

\bibitem[{\citenamefont{O'Brien et~al.}(2004)\citenamefont{O'Brien, McPeake,
  Ramakrishna, and Pendry}}]{OMR4}
\bibinfo{author}{\bibfnamefont{S.}~\bibnamefont{O'Brien}},
  \bibinfo{author}{\bibfnamefont{D.}~\bibnamefont{McPeake}},
  \bibinfo{author}{\bibfnamefont{S.~A.} \bibnamefont{Ramakrishna}},
  \bibnamefont{and} \bibinfo{author}{\bibfnamefont{J.~B.}
  \bibnamefont{Pendry}}, \bibinfo{journal}{Phys.\ Rev.~B}
  \textbf{\bibinfo{volume}{69}}, \bibinfo{pages}{241101(R)}
  (\bibinfo{year}{2004}).

\bibitem[{\citenamefont{Scalora et~al.}(2005)\citenamefont{Scalora, Syrchin,
  Akozbek, Poliakov, D'Aguanno, Mattiucci, Bloemer, and Zheltikov}}]{SSA5}
\bibinfo{author}{\bibfnamefont{M.}~\bibnamefont{Scalora}},
  \bibinfo{author}{\bibfnamefont{M.~S.} \bibnamefont{Syrchin}},
  \bibinfo{author}{\bibfnamefont{N.}~\bibnamefont{Akozbek}},
  \bibinfo{author}{\bibfnamefont{E.~Y.} \bibnamefont{Poliakov}},
  \bibinfo{author}{\bibfnamefont{G.}~\bibnamefont{D'Aguanno}},
  \bibinfo{author}{\bibfnamefont{N.}~\bibnamefont{Mattiucci}},
  \bibinfo{author}{\bibfnamefont{M.~J.} \bibnamefont{Bloemer}},
  \bibnamefont{and} \bibinfo{author}{\bibfnamefont{A.~M.}
  \bibnamefont{Zheltikov}}, \bibinfo{journal}{Phys.\ Rev.\ Lett.}
  \textbf{\bibinfo{volume}{95}}, \bibinfo{pages}{013902}
  (\bibinfo{year}{2005}).

\bibitem[{\citenamefont{Popov and Shalaev}(2006)}]{PopSha6}
\bibinfo{author}{\bibfnamefont{A.~K.} \bibnamefont{Popov}} \bibnamefont{and}
  \bibinfo{author}{\bibfnamefont{V.~M.} \bibnamefont{Shalaev}},
  \bibinfo{journal}{Opt. Lett.} \textbf{\bibinfo{volume}{31}},
  \bibinfo{pages}{2169} (\bibinfo{year}{2006}).

\bibitem[{\citenamefont{Padilla et~al.}(2006)\citenamefont{Padilla, Taylor,
  Highstrete, Lee, and Averitt}}]{PadTayHig6}
\bibinfo{author}{\bibfnamefont{W.~J.} \bibnamefont{Padilla}},
  \bibinfo{author}{\bibfnamefont{A.~J.} \bibnamefont{Taylor}},
  \bibinfo{author}{\bibfnamefont{C.}~\bibnamefont{Highstrete}},
  \bibinfo{author}{\bibfnamefont{M.}~\bibnamefont{Lee}}, \bibnamefont{and}
  \bibinfo{author}{\bibfnamefont{R.~D.} \bibnamefont{Averitt}},
  \bibinfo{journal}{Phys.\ Rev.\ Lett.} \textbf{\bibinfo{volume}{96}},
  \bibinfo{pages}{107401} (\bibinfo{year}{2006}).

\bibitem[{\citenamefont{Gabitov et~al.}(2006)\citenamefont{Gabitov, Indik,
  Litchinitser, Maimistov, Shalaev, and Soneson}}]{GabIndLit6}
\bibinfo{author}{\bibfnamefont{I.~R.} \bibnamefont{Gabitov}},
  \bibinfo{author}{\bibfnamefont{R.~A.} \bibnamefont{Indik}},
  \bibinfo{author}{\bibfnamefont{N.~M.} \bibnamefont{Litchinitser}},
  \bibinfo{author}{\bibfnamefont{A.~I.} \bibnamefont{Maimistov}},
  \bibinfo{author}{\bibfnamefont{V.~M.} \bibnamefont{Shalaev}},
  \bibnamefont{and} \bibinfo{author}{\bibfnamefont{J.~E.}
  \bibnamefont{Soneson}}, \bibinfo{journal}{J.~Opt.\ Soc.\ Am.~B}
  \textbf{\bibinfo{volume}{23}}, \bibinfo{pages}{535} (\bibinfo{year}{2006}).

\bibitem[{\citenamefont{Syms et~al.}(2008)\citenamefont{Syms, Solymar, and
  Young}}]{SymSolYou8}
\bibinfo{author}{\bibfnamefont{R.~R.~A.} \bibnamefont{Syms}},
  \bibinfo{author}{\bibfnamefont{L.}~\bibnamefont{Solymar}}, \bibnamefont{and}
  \bibinfo{author}{\bibfnamefont{I.~R.} \bibnamefont{Young}},
  \bibinfo{journal}{Metamaterials} \textbf{\bibinfo{volume}{2}},
  \bibinfo{pages}{122} (\bibinfo{year}{2008}).

\bibitem[{\citenamefont{Poutrina et~al.}(2010)\citenamefont{Poutrina, Huang,
  and Smith}}]{PouHuaSmi10}
\bibinfo{author}{\bibfnamefont{E.}~\bibnamefont{Poutrina}},
  \bibinfo{author}{\bibfnamefont{D.}~\bibnamefont{Huang}}, \bibnamefont{and}
  \bibinfo{author}{\bibfnamefont{D.~R.} \bibnamefont{Smith}},
  \bibinfo{journal}{New J.\ Phys.} \textbf{\bibinfo{volume}{12}},
  \bibinfo{pages}{093010} (\bibinfo{year}{2010}).

\bibitem[{\citenamefont{Boardman et~al.}(2011)\citenamefont{Boardman,
  Grimalsky, Kivshar, Koshevaya, Lapine, Litchinitser, Malnev, Noginov,
  Rapoport, and Shalaev}}]{BoaGriKiv11}
\bibinfo{author}{\bibfnamefont{A.}~\bibnamefont{Boardman}},
  \bibinfo{author}{\bibfnamefont{V.}~\bibnamefont{Grimalsky}},
  \bibinfo{author}{\bibfnamefont{Yu.}~\bibnamefont{Kivshar}},
  \bibinfo{author}{\bibfnamefont{S.}~\bibnamefont{Koshevaya}},
  \bibinfo{author}{\bibfnamefont{M.}~\bibnamefont{Lapine}},
  \bibinfo{author}{\bibfnamefont{N.}~\bibnamefont{Litchinitser}},
  \bibinfo{author}{\bibfnamefont{V.}~\bibnamefont{Malnev}},
  \bibinfo{author}{\bibfnamefont{M.}~\bibnamefont{Noginov}},
  \bibinfo{author}{\bibfnamefont{Y.}~\bibnamefont{Rapoport}}, \bibnamefont{and}
  \bibinfo{author}{\bibfnamefont{V.~M.} \bibnamefont{Shalaev}},
  \bibinfo{journal}{Laser Photonics Rev.}
  \textbf{5}, \bibinfo{pages}{287} (\bibinfo{year}{2011}).

\bibitem[{\citenamefont{Lapine et~al.}(2009)\citenamefont{Lapine, Gorkunov,
  Powell, Shadrivov, Marqu{\'e}s, and Kivshar}}]{LapPowGor09}
\bibinfo{author}{\bibfnamefont{M.}~\bibnamefont{Lapine}},
  \bibinfo{author}{\bibfnamefont{M.}~\bibnamefont{Gorkunov}},
  \bibinfo{author}{\bibfnamefont{D.~A.} \bibnamefont{Powell}},
  \bibinfo{author}{\bibfnamefont{I.~V.} \bibnamefont{Shadrivov}},
  \bibinfo{author}{\bibfnamefont{R.}~\bibnamefont{Marqu{\'e}s}},
  \bibnamefont{and} \bibinfo{author}{\bibfnamefont{Yu.~S.}
  \bibnamefont{Kivshar}}, \bibinfo{journal}{Appl.\ Phys.\ Lett.}
  \textbf{\bibinfo{volume}{95}}, \bibinfo{pages}{084105}
  (\bibinfo{year}{2009}).

\bibitem[{\citenamefont{Powell et~al.}(2010)\citenamefont{Powell, Lapine,
  Gorkunov, Shadrivov, and Kivshar}}]{PowLapGor10}
\bibinfo{author}{\bibfnamefont{D.~A.} \bibnamefont{Powell}},
  \bibinfo{author}{\bibfnamefont{M.}~\bibnamefont{Lapine}},
  \bibinfo{author}{\bibfnamefont{M.}~\bibnamefont{Gorkunov}},
  \bibinfo{author}{\bibfnamefont{I.~V.} \bibnamefont{Shadrivov}},
  \bibnamefont{and} \bibinfo{author}{\bibfnamefont{Yu.~S.}
  \bibnamefont{Kivshar}}, \bibinfo{journal}{Phys.\ Rev.~B}
  \textbf{\bibinfo{volume}{82}}, \bibinfo{pages}{155128}
  (\bibinfo{year}{2010}).

\bibitem[{\citenamefont{Gorkunov et~al.}(2002)\citenamefont{Gorkunov, Lapine,
  Shamonina, and Ringhofer}}]{GLS2}
\bibinfo{author}{\bibfnamefont{M.}~\bibnamefont{Gorkunov}},
  \bibinfo{author}{\bibfnamefont{M.}~\bibnamefont{Lapine}},
  \bibinfo{author}{\bibfnamefont{E.}~\bibnamefont{Shamonina}},
  \bibnamefont{and} \bibinfo{author}{\bibfnamefont{K.~H.}
  \bibnamefont{Ringhofer}}, \bibinfo{journal}{Eur.\ Phys.\ J.~B}
  \textbf{\bibinfo{volume}{28}}, \bibinfo{pages}{263} (\bibinfo{year}{2002}).

\bibitem[{\citenamefont{Marqu{\'e}s et~al.}(2008)\citenamefont{Marqu{\'e}s,
  Mart{\'\i}n, and Sorolla}}]{MarMarSor}
\bibinfo{author}{\bibfnamefont{R.}~\bibnamefont{Marqu{\'e}s}},
  \bibinfo{author}{\bibfnamefont{F.}~\bibnamefont{Mart{\'\i}n}},
  \bibnamefont{and} \bibinfo{author}{\bibfnamefont{M.}~\bibnamefont{Sorolla}},
  \emph{\bibinfo{title}{Metamaterials with {N}egative {P}arameters}}
  (\bibinfo{publisher}{Wiley}, \bibinfo{year}{2008}).

\bibitem[{\citenamefont{Landau and Lifschitz}(1984)}]{Landau}
\bibinfo{author}{\bibfnamefont{L.~D.} \bibnamefont{Landau}} \bibnamefont{and}
  \bibinfo{author}{\bibfnamefont{E.~M.} \bibnamefont{Lifschitz}},
  \emph{\bibinfo{title}{Electrodynamics of {C}ontinuous {M}edia}}
  (\bibinfo{publisher}{Pergamon Press}, \bibinfo{address}{Oxford},
  \bibinfo{year}{1984}).

\end{thebibliography}
\end{document}